\documentstyle[editedvolume,epsf]{crckapb} 

\newcommand{\ts}{\thinspace}
\newcommand{\arcs}{$^{\prime\prime}$}

\begin{opening}

\title{ULTRALUMINOUS INFRARED GALAXIES}

\author{D.B. SANDERS, J.A. SURACE, AND C.M. ISHIDA}

\institute{Institute for Astronomy, University of Hawai`i\\
           2680 Woodlawn Drive, Honolulu, HI 96822, USA}

\end{opening}

\def\eps@scaling{.95}

\def\plotone#1{\centering \leavevmode
\epsfxsize=\eps@scaling\columnwidth \epsfbox{#1}}

\def\lesssim{\mathrel{\hbox{\rlap{\hbox{\lower4pt\hbox{$\sim$}}}\hbox{$<$}}}}
\def\gtrsim{\mathrel{\hbox{\rlap{\hbox{\lower4pt\hbox{$\sim$}}}\hbox{$>$}}}}

\begin{document}

\begin{abstract}
At luminosities above $\sim{\ts}10^{11}${\ts}$L_\odot$, infrared galaxies
become the dominant population of extragalactic objects in the local Universe
($z <${\ts}0.5), being more numerous than optically selected starburst and
Seyfert galaxies, and QSOs at comparable bolometric luminosity.  At the highest
luminosities, ultraluminous infrared galaxies  (ULIGs:  $L_{\rm ir} >
10^{12}{\ts}L_\odot$), \index{S:infrared galaxies!ultraluminous} outnumber
optically selected QSOs by a factor of $\sim${\ts}1.5--2{\ts}.  All of the
nearest ULIGs ($z${\ts}$<$0.1) appear to be advanced mergers that are powered
by both a circumnuclear starburst and AGN, both of which are fueled by an
enormous concentration of molecular gas ($\sim${\ts}$10^{10}${\ts}$M_\odot$)
that has been funneled into the merger nucleus. ULIGs may represent a
\index{S:gas!molecular!masses} \index{S:gas!molecular!nuclear} primary stage in
the formation of massive black holes and elliptical galaxy
\index{S:quasars!triggering} cores. The intense circumnuclear starburst that
accompanies the ULIG phase may also represent a primary stage in the formation
of globular clusters, and the metal enrichment of the intergalactic medium by
gas and dust expelled from the nucleus due to the combined forces of supernova
explosions and powerful stellar winds.  
\end{abstract}

\vspace{-0.4cm}
\section{Introduction}

\begin{figure}[htbp]
\plotone{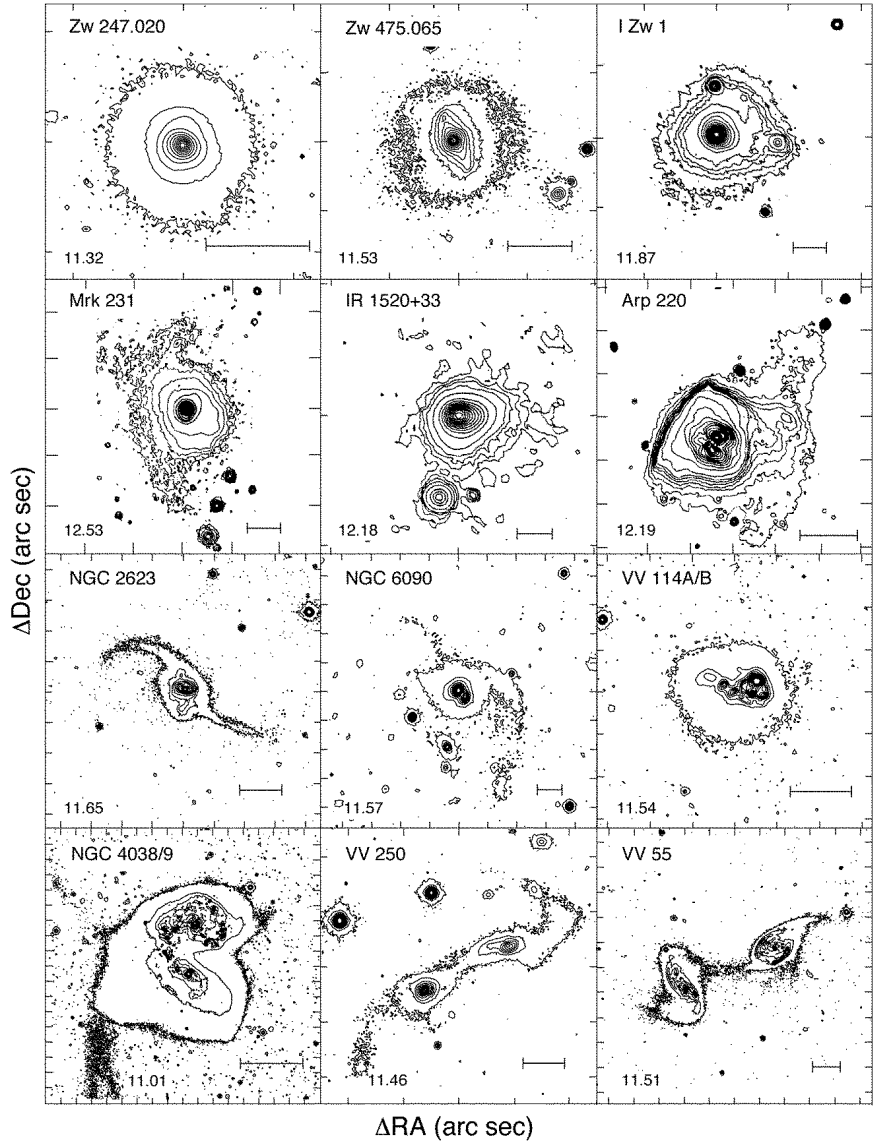}
\caption{R-band images of a subset of 12 LIGs selected from the 
{\it IRAS} Revised Bright Galaxy Sample (RBGS: Sanders et al. 1998) 
and a complete sample of ``warm'' ULIGs (Sanders et al. 1988b).  
The scale bar represents 10{\ts}kpc, tick marks are at 20\arcs\ 
intervals, and the infrared luminosity (log{\ts}$L_{\rm ir}/L_\odot$) 
is indicated in the lower left corner of each panel. This subsample is 
chosen to illustrate the full range of morphologies and infrared 
luminosities found in the complete sample of LIGs and `warm' ULIGs  -- 
from the most luminous ULIGs which appear to contain dominant single 
nuclei (e.g. Mrk~231, I~Zw~1), to lower luminosity sources that are 
either pairs of distinct, tidally distorted disks in the 
early stage of merger (bottom row), or apparently single objects with 
elliptical-like radial light profiles that may be the most advanced 
and relaxed mergers (e.g. Zw~247.020, Zw~475.056).  These ground-based data 
(typical seeing is 0.7\arcs--1.2\arcs) are currently being replaced with 
higher resolution {\it HST} and ground-based adaptive optics images at 
UV-to-nearIR wavelengths.  As an example, see \index{P:5} \index{P:6}  
color plates~5-6 (pp.~xxiii-xxiv) for new data on NGC 4038/39.}
\end{figure}

One of the major results of the Infrared Astronomical Satellite ({\it IRAS})
\index{S:spacecraft!IRAS@{\it IRAS}} all-sky survey \index{S:surveys!{\it IRAS}
All-Sky} was the identification of a class of luminous infrared galaxies
(LIGs:  $L_{\rm ir} > 10^{11}\ L_\odot$;  $H_{\rm o} = 75${\ts}km
s$^{-1}$Mpc$^{-1}$, $q_{\rm o} = 0.5$)\footnote{$L_{\rm ir} \equiv
L$(8-1000$\mu$m), computed from the observed infrared fluxes in all four {\it
IRAS\/} bands according to the prescription in Perault (1987); see also Sanders
\& Mirabel (1996).}, objects that emit more energy in the
far-infrared/submillimeter than at all other wavelengths combined.  Redshift
surveys of complete samples of {\it IRAS} galaxies now agree that infrared
selected galaxies become the dominant population of extragalactic objects at
bolometric luminosities above $\sim{\ts}4{\ts}L^*$ (i.e. $L_{\rm bol}
>{\ts}10^{11}{\ts}L_\odot$).  Reasonable assumptions about the lifetime of the
infrared phase suggest that a {\it substantial fraction of all galaxies with}
$L_{\rm B} >{\ts}10^{10}{\ts}L_\odot$ {\it may at some point in their lifetime
pass through such a stage of intense infrared emission (Soifer et al 1987)}.
This review focuses on providing an up-to-date summary of the observed
morphological properties of ULIGs\footnote{Optical/near-IR spectroscopy of
LIGs, and the nuclear gas and dust properties of ULIGs are discussed in
conference papers by Veilleux and Scoville \& Yun respectively.}, those
infrared-selected objects which represent an extreme phase of nuclear activity
in galaxies, equivalent to the bolometric luminosity of optically selected
QSOs.

\section{LIGs and ULIGs: A Merger Sequence} 
Ground-based optical and
near-infrared imaging of complete samples of the brightest infrared galaxies
clearly show that a substantial fraction of LIGs are strongly interacting or
merging spirals, and that the higher the luminosity the more advanced is the
merger.  Millimeterwave observations of have shown these spirals to be rich in
molecular gas  -- $M({\rm H}_2) \sim 10^9-3\times 10^{10}${\ts}$M_\odot$ (e.g.
Sanders et al. 1991) -- and that there is an increasing central concentration
of this gas with increasing infrared luminosity.  The representative subsample
of LIGs shown in Figure 1 illustrates the signs of strong interactions/mergers
(tidal tails, double nuclei, etc.) that are revealed in deep optical images of
nearby LIGs.  Comparison of the images with numerical simulations (e.g. Barnes
\& Hernquist 1992; Mihos \& Hernquist 1994) allows these objects to be placed
in a rough time sequence.

\section{ULIG Properties}

\begin{figure}[htbp]
\vspace{-0.3cm}
\plotone{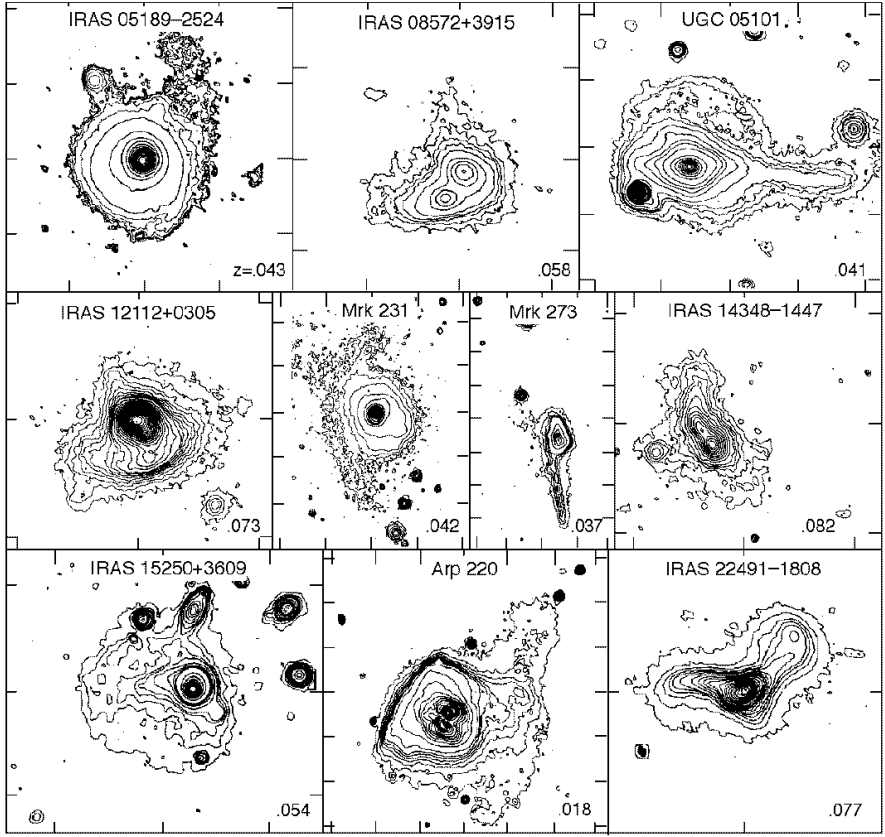}
\vspace{-0.2cm}
\caption{Optical (r-band) CCD images of the complete sample of 10 ULIGs 
from the original {\it IRAS} BGS (Sanders et al. 1988a).  Tick marks 
are at 20\arcs\ intervals.  Typical seeing for these ground-based images 
is $\sim$0.8\arcs--1.5\arcs.}
\vspace{-0.5cm}
\end{figure}

\begin{table}[hbp]
\caption{Properties of ULIGs}
\begin{center}
\begin{tabular}{lccc}
Property &  Median & Min & Max \\
\hline
redshift & 0.05 & 0.018 & 0.136 \\
log{\ts}$L_{\rm ir}$\ [$L_\odot$]& 12.2 & 12.0 & 12.65 \\
log{\ts}$M({\rm H}_2)$\ [$M_\odot$] & 10.0 & 9.3 & 10.7 \\
$M({\rm H}_2)$\ at $r<$1kpc\ [\%] & 65 & 40 & 100 \\
$<\sigma({\rm H}_2)>$ at $r<$0.5kpc\ [$M_\odot pc^{-2}$] & 
  $4 \times 10^4$  & $1 \times 10^4$ & $1 \times 10^5$ \\
$<\rho({\rm H}_2)>$ at $r<$0.5kpc\ [$M_\odot pc^{-3}$] & 
  $3 \times 10^2$  & $1 \times 10^2$ & $1 \times 10^3$ \\
$<A_{\rm V}>$ towards nucleus\ [mags] & 800 & 400 & 2000  \\
nuclear separation\ [kpc] & 1.9 & $<$0.02 & 9.3 \\
tail length\ [kpc] & 45  & 20  & 120  \\
B-band luminosity\ [$L^*_{\rm B}$] & 2.5 & 1.1 & 4.4\\
K-band luminosity\ [$L^*_{\rm K}$] & 2.5 & 1.2 & 7.3 \\
\hline
\end{tabular}
\end{center}
\end{table}

Nearly all ULIGs appear to be late-stage mergers \index{S:mergers!advanced}
(e.g. Sanders et al 1988a,b; Melnick \& Mirabel 1990; Kim 1995; Murphy et al.
1996; Clements et al. 1996).  Figure 2 illustrates the largely overlapping
disks \index{S:mergers!disk galaxies}that are seen \index{S:mergers!signatures}
in a {\it complete} sample of the nearest and best-studied ULIGs.  The true
extent of faint tidal features plus greater detail in the inner disks of these
ULIGs is better revealed in the higher resolution ground-based images and {\it
HST} images of ULIGs shown in \index{P:7} \index{P:8} color plates~7-8
(pp.~xxv-xxiv).  Table~1 summarizes properties of the complete sample of 20
ULIGs from the {\it IRAS} Bright Galaxy Samples (Soifer et al 1987; Sanders et
al 1995).  The mean lifetime for the ULIG phase, estimated from the observed
mean separation and relative velocity of the merger nuclei, is
$\sim${\ts}$2-4\times 10^8${\ts}yrs.  \index{S:mergers!timescale}

\vspace{-0.4cm}
\section{The Nuclear Starburst-AGN Connection and the Fate of ULIGs}

\begin{figure}[htbp]
\plotone{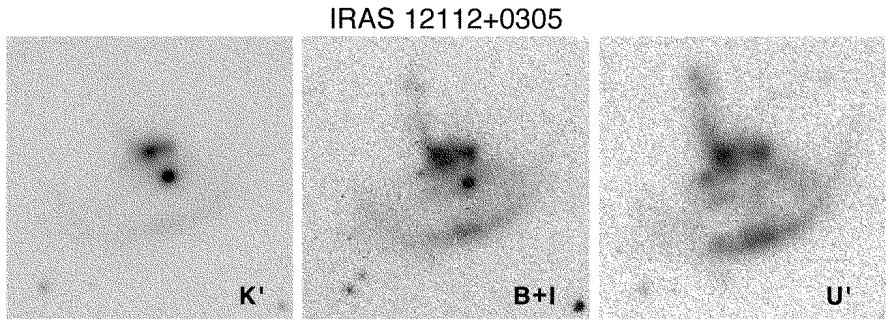}
\caption{UV/Optical and near-infrared ground-based images of IRAS 12112+0305
(Surace 1998).  The short wavelengths are dominated by knotty star formation,
both in the central regions and along the extended tidal features.  However
most of the bolometric luminosity of this system appears to originate in the
central knot ($d <$200{\ts}pc) that is completely obscured at U$^\prime$, and
that is either powered by an AGN, a superstarburst (which by itself would be
much more powerful than the sum of the luminosity from all other starburst
regions in this object), or a mixture of starburst and AGN.} 
\end{figure}

The enormous central gas supplies present in ULIGs are clearly an 
ideal breeding ground 
for both powerful circumnuclear starbursts and AGN.  Indeed those 
ULIGs that have been imaged with adaptive optics from the ground 
(see Fig. 3) or with {\it HST} (see Fig. 4) show evidence for a 
population of massive \index{S:stellar populations!young} young 
($\sim${\ts}10$^7${\ts}yrs) star clusters, although these 
clusters account for typically much less that half the ULIG 
bolometric luminosity (Surace \& Sanders 1999).  
Most of the luminosity appears to be 
concentrated in one or two small ($r <$100{\ts}pc) regions 
centered on the putative nucleus (or nuclei) (e.g. Soifer, et al. 1998), 
and it is these compact regions which most likely harbor exotic 
superstarbursts, \index{S:starbursts!nuclear} and/or a powerful 
AGN\index{S:AGN}. 

There is now substantial evidence that ULIGs are elliptical 
galaxies forming by merger-induced \index{S:mergers!dissipative} 
\index{S:mergers!elliptical galaxies} \index{S:mergers!remnants} 
dissipative collapse (Kormendy \& Sanders 1992), including $r^{1/4}$-law 
brightness profiles (e.g. Schweizer 1982; Wright et al. 1990; Kim 1995), 
newly-formed globular clusters \index{S:globular clusters!formation} 
(e.g. Fig. 4 and Surace et al. 1998),  
central gas densities ($\gtrsim 10^2 M_\odot pc^{-3}$ at 
$r \lesssim 0.5-1${\ts}kpc: Scoville et al. 1991) that are as high 
\index{S:gas!molecular!nuclear}
as stellar mass densities in the cores of giant ellipticals, and powerful 
``superwinds" (Heckman et al. 1987; Armus et al. 1989) which will 
likely leave behind a largely dust free core.  It seems reasonable 
to assume also, that this scenario might lead to a constant 
ratio of black hole mass to bulge mass in agreement with recent 
observational results (Kormendy \& Richstone 1995; Magorrian et al. 1998). 

Future infrared space missions and more sensitive 
submillimeter surveys should succeed in identifying more distant 
ULIGs, thus allowing a direct test of whether the infrared 
luminosity function evolves as steeply as that of QSOs, 
and whether ULIGs were more numerous at $z \sim 1-4$ when it 
is presumed that most of the ellipticals were formed from 
mergers of spirals.

\begin{figure}[htbp]
\plotone{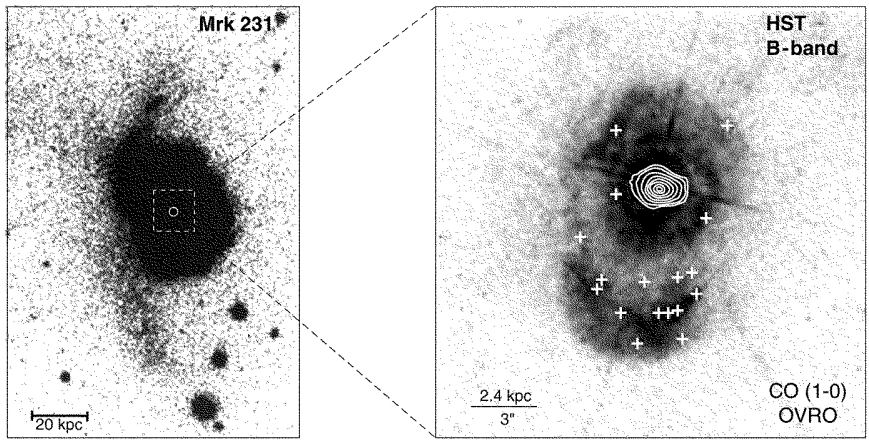}
\caption{The advanced merger/ULIG/QSO Mrk 231 -- Left panel: optical image
(Sanders et al. 1987) and CO contour (Scoville et al. 1989).\ Right panel:
{\it HST} B-band image and identified stellar clusters (`+') from Surace et al.
(1998).  The high resolution CO contours are from Bryant \& Scoville (1996)}
\vspace{-0.4cm} 
\end{figure}

\end{document}